\documentclass[vecphys]{svmult}
\usepackage{makeidx, multicol, amsmath, amssymb, enumerate, bbm}
\usepackage[bottom]{footmisc}

\begin{document}
 
\title*{Non-extremal black holes from the generalised r-map}

\author{Thomas Mohaupt and Owen Vaughan}

\institute{Department of Mathematical Sciences, 
University of Liverpool, Peach Street, 
Liverpool L69 7ZL, UK.  \\
E-mail: \texttt{thomas.mohaupt@liv.ac.uk}, \texttt{owen.vaughan@liv.ac.uk}}

\maketitle

\vspace{-17em}
\rightline{LTH 956} 
\vspace{17em}

\section*{Abstract}

\begin{quotation} \noindent
We review the timelike dimensional reduction of a class of five-dimensional theories that generalises $5D, \, {\cal N} = 2$ supergravity coupled to vector multiplets. As an application we construct instanton solutions to the four-dimensional Euclidean theory, and investigate the criteria for solutions to lift to static non-extremal black holes in five dimensions. 

We focus specifically on two classes of models: STU-like models, and models with a block diagonal target space metric. For STU-like models the second order equations of motion of the four-dimensional theory can be solved explicitly, and we obtain the general solution. For block diagonal models we find a restricted class of solutions, where the number of independent scalar fields depends on the number of blocks. When lifting these solutions to five dimensions we show, by explicit calculation, that one obtains static non-extremal black holes with scalar fields that take finite values on the horizon only if the number of integration constants reduces by exactly half.

\end{quotation}


\section{Introduction}

Black holes provide an important testing ground for string theory and other theories of quantum gravity. Theories with extended supersymmetry allow for extremal BPS black hole solutions, and for certain examples the microscopic and macroscopic entropy has been calculated with agreement to leading order \cite{Ferrara:1995ih, Strominger:1996sh}, and even to higher orders when including $R^2$ corrections \cite{Maldacena:1997de, Vafa:1997gr, Lopes Cardoso:1998wt}. Interestingly, the entropy of certain near-extremal black holes can also be calculated \cite{Callan:1996dv, Horowitz:1996fn, Cvetic:1996kv, Maldacena:1996ix, Behrndt:1997as}, with at least leading order agreement. In order to improve on this analysis it is critical to have a systematic understanding of non-extremal black hole solutions of lower-dimensional supergravity theories. 
This naturally leads one to consider maps between the various special geometries of ${\cal N} = 2$ supergravity through dimensional reduction, which are also interesting mathematically. These go by the names of the r-map and c-map, and although they have been known for some time \cite{Cecotti:1988qn, Ferrara:1989ik, deWit:1991nm, deWit:1995tf}, they have also seen much recent interest, a small sample of which is given by \cite{Gunaydin:2005mx, Bossard:2009at, Chemissany:2009hq, Bossard:2009we, AzregAinou:2011gt, Cortes:2011, Mohaupt:2011aa, Alexandrov:2012bn, Alekseevsky:2012}. Dimensional reduction over time need not be restricted to supersymmetric theories \cite{Yazadjiev:2005pf, Gaiotto:2007ag, Janssen:2007rc, Perz:2008kh}, with the standard reference for non-linear sigma models coupled to vector fields and gravity being the seminal paper \cite{Breitenlohner:1987dg}.
Since static, single-centred black hole solutions correspond to geodesics in the target manifold of the image of these maps, there exists a rich interplay between physical objects and geometrical constructions.

We will review the procedure presented in \cite{Mohaupt:2010fk} for producing non-extremal static black hole solutions to a large class of five-dimensional theories, which includes ${\cal N} = 2$ supergravity coupled to vector multiplets as a subclass. The method is based on \cite{Mohaupt:2009iq}, and uses dimensional reduction (the r-map) over a timelike direction followed by a specific field redefinition, which can be understood as follows:
The physical scalar fields parametrise a hypersurface in a larger ambient space (a $d$-conical affine special real manifold). The field redefinition combines the physical scalar fields with the Kaluza-Klein scalar, which can be used to parametrise the direction orthogonal to the hypersurface. The new scalar fields then parametrise the whole of the ambient space. 
After this procedure the effective Lagrangian for static, spherically symmetric and purely electric backgrounds takes the particularly simple form:
	\[
		\mathtt{e}^{-1}_4 {\cal L}_4 = \tfrac12 R_4 - \tfrac{3}{4}a_{IJ}(\sigma) \left( \partial_\mu \sigma^I \partial^\mu \sigma^J  - \partial_\mu b^I \partial^\mu b^J \right)  \;. 
	\]
Here $\sigma^I$ are the scalars fields which combine the original five-dimensional physical scalars with the Kaluza-Klein scalar. The axionic scalar fields $b^I$ descend from the gauge sector, and represent the electric potentials.

Solving the equations of motion corresponds to constructing harmonic maps from reduced spacetime into a target manifold, which becomes enlarged due to the dimensional reduction procedure. We focus on STU-like models, for which the general solution to the full second order equations of motion can be found. This is a class of models that contains the STU model along with specific generalisations that share the same feature of having a diagonal target space metric. We also consider models with block diagonal target space metrics, where a restricted class of solutions can be found that is based on the solutions to STU-like models. We will see that the number of independent scalar fields in these solutions depends on the number of blocks in the metric. For all models this provides one universal solution with constant scalar fields, because all metrics can be thought of as having at least one block (the whole metric). 

We then investigate the criteria for solutions to correspond to static, non-extremal black holes in the five-dimensional theory with scalar fields that take finite values on the horizon. We find that the number of integration constants must reduce by half, which is suggestive of a first order rewriting. While first order equations governing examples of non-extremal black holes have been known for some time \cite{Lu:2003iv, Miller:2006ay, Garousi:2007zb, Andrianopoli:2007gt, Janssen:2007rc, Cardoso:2008gm, Perz:2008kh, Chemissany:2009hq}, it has previously been used (to our knowledge) only as an ansatz for obtaining specific non-extremal solutions. The logic presented here is different. We consider the most general type of solution and then restrict it to solutions that describe non-extremal black holes. For STU-like models all calculations are performed explicitly, and actually rather simply.

Since this method does not rely critically on supersymmetry, we are able to consider a larger class of theories than  $5D,\,{\cal N} = 2$ supergravity coupled to vector multiplets. This is achieved by generalising the geometry of the target manifold of the scalar fields in two ways: firstly, we do not require that the Hesse potential (often called the prepotential) is a homogeneous polynomial, but just a homogeneous function. Secondly, we allow the degree of homogeneity not just to be three, but to be arbitrary. Mathematically, this means that we replace the projective special real target manifold, which is required for $5D,\,{\cal N} = 2$ supergravity \cite{Gunaydin:1983bi}, with a generalised projective special real manifold. The generalisation is captured in the degree of homogeneity of the Hesse potential of the corresponding $d$-conical affine special real manifold \cite{Cortes:2011aj}. The kinetic terms of the gauge fields also get modified in an appropriate fashion. We refer to the dimensional reduction of such a theory as the generalised r-map. Various geometrical aspects of this map have been discussed in \cite{Cortes:2011aj}, and the analogous generalisation of the rigid r-map has been also been considered in \cite{AlekseevskyRigid}.

\subsection{The Reissner-Nordstr\"om black hole}

Let us first briefly review the five-dimensional Reissner-Nordstr\"om black hole, which will be our guiding example. This is a static, spherically symmetric and purely electric solution to a five-dimensional theory of gravity coupled to a single $\text{U}(1)$ gauge field. The line element for this solution can be written as
\begin{equation}
	ds_{5}^2 = -\frac{W}{ {\cal H}^2 } dt^2 + {\cal H} \left( W^{-1} d r^2 + r^2 d\Omega^2_{(3)} \right) \;,
	\label{5dRN}
\end{equation}
where the functions ${\cal H}$ and $W$ are given by
\[
	{\cal H} = 1 + \frac{q}{r^2} \;, \qquad W = 1 - \frac{2c}{r^2} \;,
\]
and are harmonic functions with respect to the flat metric on ${\mathbb R}^4$, i.e. 
\[
	\Delta_{4} {\cal H} = \Delta_{4} W = 0 \;.
\]
The parameter $q$ is the electric charge, and $c$ is the non-extremality parameter. The mass is given by $m^2 = q^2 - c^2$.
In these coordinates the solution has an outer event horizon at $r = 2c$ and an inner Cauchy horizon at $r = 0$. One can analytically continue these coordinates to the singularity, which is located at $r = - \sqrt{q}$. The extremal limit in given by $c \to 0$, in which case $W \to 1$. 
It will be useful later to decompose the five-dimensional metric according to
\[
	ds^2_5 = -e^{2\bar{\sigma}\phi} dt^2 + e^{-\bar{\sigma}} ds^2_4 \;,
\]
which for the Reissner-Nordstr\"om metric corresponds to
\begin{equation}
	e^{\bar{\sigma}} = \frac{\sqrt{ W}}{{\cal H}} \;, \qquad ds^2_4 = \frac{d r^2}{\sqrt{W}} + \sqrt{W} r^2 d\Omega^2_{(3)} \;.
	\label{RN_decomposition}
\end{equation}

The simple example of the Reissner-Nordstr\"om black hole gives us some important clues about non-extremal solutions:
\begin{enumerate}[(i)]
	\item The solution is built from harmonic functions on ${\mathbb R}^4$. 
	\item The four-dimensional line element is flat in the extremal limit.
	\item The non-extremal solution is obtained by dressing the extremal solution with one additional harmonic function $W$.
\end{enumerate}
We will see that these key features of the Reissner-Nordstr\"om black hole are also true of more complicated non-extremal solutions.


\section{Generalising $5D, \, {\cal N} = 2$ supergravity}

Before we write down the Lagrangian of the class of theories under consideration, we will first give a mathematical description of generalised projective special real geometry, which is a generalisation of the geometry of $5D$ vector multiplets. This is based in part on \cite{Cortes:2011aj}, work in progress with Vicente Cort\'es and the first author, and a summary given in \cite{Mohaupt:2011ab}. The less mathematically inclined reader may skip this section and move directly to section 2.2.

\subsection{Generalising special real geometry}

A $d$-conical affine special real manifold $(M, g, \nabla, \xi)$ is a pseudo-Riemannian manifold $(M,g)$ of $\text{dim}_{\mathbb R}M = (n + 1)$ equipped with a flat, torsion free `special' connection $\nabla$ and vector field $\xi$ such that 
\begin{enumerate}[(i)]
	\item $\nabla g$ is completely symmetric.
	\item $D \xi = \frac{d}{2} {\mathbbm 1}$, where $D$ is the Levi-Civita connection.
	\item $\nabla \xi = {\mathbbm 1}$.
\end{enumerate}
Let us discuss each condition in turn. Firstly one can define a natural set of special coordinates $h^I$ that are flat with respect to $\nabla$, i.e.
\[	
	\nabla d h^I =  0\;, \qquad \Rightarrow \qquad \nabla_I = \partial_I \;.
\]
With respect to these coordinates the condition (i) ensures that
\[
	\frac{\partial}{\partial h^I} g_{JK}(h) = \frac{\partial}{\partial h^J} g_{IK}(h) \;,
\]
and, hence, the metric $g$ is given by the second derivatives of a function
\[
 g = \partial^2 H \;.
\]
Such a function is referred to a Hesse potential, and it is not unique.
For condition (ii) we follow a similar analysis to \cite{Gibbons:1998xa}, which deals with the particular case $d = 2$. This condition implies that $\xi$ is a homothetic Killing vector field of weight $d$
\[
	{\cal L}_\xi g = dg \;.
\]
Moreover it ensures that the manifold has the property of being $d$-conical, which means there always exists a coordinate system $(r,x^i)$, with $r^d = g(\xi,\xi)$, such that the metric decomposes as 
\[
	g = r^{d-2} dr^2 + r^d \bar{g}(x^i) \;. 
\]
In these coordinates $\xi = r\frac{\partial}{\partial r}$. One can then define the new coordinates $y^I = (r, rx^i)$, for which the homothetic Killing vector $\xi$ becomes an Euler vector field
\[
	\xi = y^I \frac{\partial}{\partial y^I} \;.
\]
In such coordinates the metric components are homogeneous functions of degree $(d - 2)$
\[
	\xi g_{IJ}(y) = (d - 2) g_{IJ}(y) \;,
\]
which can be deduced from the fact that $\left[\xi, \frac{\partial}{\partial y^I} \right] = -\frac{\partial}{\partial y^I}$.
The last condition (iii) can be seen as a compatibility condition between the previous two conditions. It ensures that $\xi$ is the Euler field associated with the special coordinates
\[	
	\xi = h^I \frac{\partial}{\partial h^I} \;,
\]
and, hence, the metric components are homogeneous functions of degree $(d - 2)$ with respect to the special coordinates $h^I$. It follows that one can always choose a unique Hesse potential that is homogeneous of degree $d$, which is given by
\[
	H = \frac{1}{d(d-1)} g_{IJ} h^I h^J \;.
\]
In order to obtain physically relevant signatures we will require this Hesse potential to be strictly positive.

It is convenient to introduce a second metric on $M$, given by
\[
	a = \partial^2 \tilde{H} \;,
\]
where $\tilde{H} := -\frac{1}{d} \log H$. We can write this metric in a basis of special coordinates as
\begin{equation}
	a = a_{IJ} dh^I \otimes dh^J = -\frac{1}{d} \left( \frac{H_{IJ}}{H} - \frac{H_I H_J}{H^2}\right) dh^I \otimes dh^J \;,
	\label{a}
\end{equation}
where $H_I, H_{IJ}$ are the first and second derivatives of the Hesse potential.
If the metric $g$ has signature $(+ - \ldots -)$, which is the case for supergravity, then $a$ is strictly positive definite. The vector field $\xi$ acts as an isometry of the metric $a$
\[
	{\cal L}_\xi a = 0 \;.
\]

We define a generalised special real manifold $(\bar{M}, \bar{g})$ as a hypersurface of constant $H$ in a $d$-conical affine special real manifold, with metric induced from $a$. If $\text{dim}_{\mathbb R} M = (n + 1)$ then $\text{dim}_{\mathbb R} \bar{M} = n$. It is particularly convenient to consider the hypersurface defined by $H = 1$ 
\[
	\bar{M} \simeq \{H = 1 \} \subset M \;,
\]
and we denote the embedding of $\bar{M}$ into $M$ given by the hypersurface $H = 1$ by $i:\bar{M} \to M$. For this embedding both the pull-back of $-\frac{1}{d}g$ and $a$ give the same metric on $\bar{M}$
\[
	\bar{g} = i^*\left( -\frac{1}{d} \partial^2 H \right) = i^*\left( \partial^2 \tilde{H} \right) \;. 
\]
Let $\phi^x$ denote local coordinates on $\bar{M}$, which therefore parametrise the hypersurface $H = 1$. The metric can be written as
\[
	\bar{g} = \bar{g}_{xy} d\phi^x \otimes d \phi^y = \left(a_{IJ} \frac{\partial h^I}{\partial \phi^x} \frac{\partial h^J}{\partial \phi^y} \right)\Bigg|_{H =1} d\phi^x \otimes d \phi^y\;.
\]
A particularly natural set of coordinates is given by 
\begin{equation}
	\phi^x = \frac{h^x}{h^0} \;, \qquad h^0 = \hat{H}(\phi^1, \ldots, \phi^n)^{-\frac{1}{d}} := H\left(1, \frac{h^1}{h^0}, \ldots, \frac{h^n}{h^0}\right)^{-\frac{1}{d}} \;. \label{Phiparam}
\end{equation}
These are analogous to the inhomogeneous special coordinates $z^i = X^i / X^0$ on a projective special K\"ahler manifold.
It is worth noting that one can also realise $\bar{M}$ as the quotient manifold $M / \mathbb{R}^{>0}$ with quotient metric obtained from $(M, a)$. 

For the special case that $d = 3$ and the Hesse potential is a polynomial then  $(\bar{M}, \bar{g})$ represents the target manifold of $5D, \,{\cal N} = 2$ supergravity coupled to vector multiplets \cite{Gunaydin:1983bi}. The matrix $a_{IJ}$ restricted to the hypersurface $H = 1$ provides the kinetic term for the gauge fields. We will make the same identifications when considering more general Lagrangians, only we no longer require that $d = 3$ or the Hesse potential is a polynomial.

\subsection{Generalising the Lagrangian}

We are now ready to generalise the Lagrangian of five-dimensional ${\cal N} = 2$ supergravity coupled to $n$ abelian vector multiplets. Our starting point is the Lagrangian of a five-dimensional theory of gravity coupled to $n$ scalar fields and $(n + 1)$ abelian gauge fields
\begin{equation}
\mathtt{e}^{-1}_5 {\cal L}_5 = \tfrac{1}{2} R_5 - \tfrac{3}{4} \bar{g}_{xy}	\partial_{\hat{\mu}} \phi^x \partial^{\hat{\mu}} \phi^y 
- \tfrac{1}{4} a_{IJ} F^I_{\hat{\mu} \hat{\nu}} F^{J \hat{\mu} \hat{\nu}} \;,
\label{GenLag}
\end{equation}
We could also have included a Chern-Simons term, however this will not be relevant for solutions which are static and purely electric. Likewise for fermionic terms. Spacetime indices run from $\hat{\mu} = 0, \ldots, 4$, and target space indices from $x = 1, \ldots n$, $I = 0, \ldots, n$. The coupling matrices $\bar{g}_{xy}$ and $a_{IJ}$ depend only on $\phi^x$. 

The scalar fields form a  non-linear sigma model with values in an $n$-dimensional target manifold that we require to be generalised projective special real (as defined in the previous section). The matrix $a_{IJ}$ are the components of the tensor field (\ref{a}) on the corresponding $d$-conical affine special real manifold. We will require that $g_{IJ}$ has signature $(+ - \ldots -)$, and, hence, $a_{IJ}$ is positive definite.
One obtains a projective special real manifold, and therefore $5D, \, {\cal N} = 2$ supergravity, for the special case when $d = 3$ and the Hesse potential is a polynomial.

We prefer not to work with the $n$ physical scalar fields $\phi^x$ but rather the $(n + 1)$ special coordinates $h^I$, which are subject to the hyper-surface constraint
\begin{equation}
	H(h) = 1 \;.
	\label{HypConst}
\end{equation}
Here $H$ is a smooth homogeneous function of degree $d$, and represents the Hesse potential of the corresponding $d$-conical affine special real manifold. It is often convenient to choose the parametrisation given by (\ref{Phiparam}), where $\phi^x$ and $h^I$ can be related explicitly.
In the Lagrangian one must make the replacement
\[
	\bar{g}_{xy}(\phi)	\partial_{\hat{\mu}} \phi^x \partial^{\hat{\mu}} \phi^y \to 
	a_{IJ}(h) \partial_{\hat{\mu}} h^I \partial^{\hat{\mu}} h^J \Big|_{H=1}\;,
\]
and, hence, the Lagrangian can be written as
\begin{equation}
\mathtt{e}^{-1}_5 {\cal L}_5 = \tfrac{1}{2} R_5 - \tfrac{3}{4} a_{IJ}	\partial_{\hat{\mu}} h^I \partial^{\hat{\mu}} h^J 
- \tfrac{1}{4} a_{IJ} F^I_{\hat{\mu} \hat{\nu}} F^{J \hat{\mu} \hat{\nu}} \;,
\label{5d_action}
\end{equation}
where it is understood that the scalar fields $h^I$ are now subject to the constraint (\ref{HypConst}).
Two advantages of using the special coordinates $h^I$ are immediately clear: we now have the same number of scalar fields as gauge fields, and the coupling matrices are the same.
The coupling matrix $a_{IJ}$ can be written in these coordinates as
\[
a_{IJ}(h) = \partial^2_{I,J} \tilde{H} (h) \;,
\]
where as in the previous section $\tilde{H} := -\frac{1}{d} \log H$. The details of the model are completely determined by the Hesse potential $H$.


\section{Dimensional reduction and equations of motion}

We now impose that backgrounds are static, and make the following decomposition of the five-dimensional metric:
\[
	ds^2_5 = -e^{2\bar{\sigma}} dt^2 + e^{-\bar{\sigma}} ds^2_4 \;,
\]
We impose further that backgrounds are purely electric, so the gauge vector and field strength decompose as
\[
	\mathcal{A}^I = \sqrt{\frac32} \, b^I dt + C^I_\mu dx^\mu \;, \qquad  C^I_\mu = \text{const} \;.
\]
Choosing $C^I_\mu$ to be constant ensures that the magnetic components of the field strengths $F^I_{\hat{\mu} \hat{\nu}}$ vanish, and we can write
\[
	F^I_{\hat{\mu} \hat{\nu}} F^{J \hat{\mu} \hat{\nu}} = - 3 e^{-2 \tilde{\sigma}} \partial_\mu b^I \partial^\mu b^J \;.
\]
The scalar fields $b^I$ represent the electric potentials.
After integrating out the redundant timelike dimension, the four-dimensional Lagrangian takes the form
\[
\mathtt{e}^{-1}_4 {\cal L}_4 = \tfrac12 R_4 - \tfrac{3}{4} \partial_\mu \tilde{\sigma} \partial^\mu \tilde{\sigma}
- \tfrac{3}{4}a_{IJ}(h) \left( \partial_\mu h^I \partial^\mu h^J - e^{-2\tilde{\sigma}}\partial_\mu b^I \partial^\mu b^J \right)  \;. 
\]
We now combine the KK-scalar $\tilde{\sigma}$ and the constrained scalar fields $h^I$ into the new scalar fields $\sigma^I$
\begin{equation}
	\sigma^I := e^{\tilde{\sigma}} h^I \;. 
	\label{FieldRedef}
\end{equation}
The $(n + 1)$ scalar fields $\sigma^I$ are unconstrained, as the KK-scalar absorbs the hypersurface constraint (\ref{HypConst}), which now becomes
\[
	H(\sigma) = e^{d\tilde{\sigma}} \;.
\]
One can therefore interpret $h^I$ and the KK-scalar $\tilde{\sigma}$ as fields that depend on $\sigma^I$, which are a set of independent fields.
Since $a_{IJ}(h)$ is homogeneous of degree $-2$ and $a_{IJ}(h)h^I \partial_\mu h^J = 0$ we have
\[
	a_{IJ}(h) \partial_\mu h^I \partial^\mu h^J = a_{IJ}(\sigma) \partial_\mu \sigma^I \partial^\mu \sigma^J 
	- \partial_\mu \tilde{\sigma} \partial^\mu \tilde{\sigma} 
	\;, 
\]
The four-dimensional Lagrangian can now be written as
\begin{equation}
	\mathtt{e}^{-1}_4 {\cal L}_4 = \tfrac12 R_4 - \tfrac{3}{4}a_{IJ}(\sigma) \left( \partial_\mu \sigma^I \partial^\mu \sigma^J  - \partial_\mu b^I \partial^\mu b^J \right)  \;. 
	\label{4dLag}
\end{equation}
This Lagrangian encodes all the information about the theory for static and purely electric backgrounds.
It will be useful later to note that the scalar fields $\sigma^I$ satisfy the relation
\begin{equation}
	a_{IJ}(\sigma) \sigma^I \sigma^J = 1 \;. \label{SigId}
\end{equation}

We will now impose that backgrounds are spherically symmetric. This is in fact enough to completely determine the four-dimensional metric%
\footnote{For a derivation see \cite{Mohaupt:2010fk}, and see \cite{Meessen:2011bd} for a general formula for $d \geq 4$ dimensions.}
\begin{equation}
	ds_{4}^2 = \frac{ c^3 }{ \sinh^3(2c \tau) } d \tau^2  + \frac{c }{ \sinh(2c \tau) } d \Omega^2_{(3)} \;. 
	\label{4D_RN_metric}
\end{equation}
Here $\tau$ is an affine parameter in the radial direction, which is related to the standard radial coordinate through
\begin{equation}
	r^2 = \frac{c e^{2c\tau}}{\sinh(2c\tau)} \;. 
	\label{eq:tau_r_relation}
\end{equation}
Subbing in $r$ to the four-dimensional metric (\ref{4D_RN_metric}) one finds that it is nothing other than the spatial part of the Reissner-Nordstr\"om metric with respect to the decomposition (\ref{RN_decomposition})
\begin{equation*}
ds^2_{4} = \frac{dr^2}{\sqrt{W}} + \sqrt{W} r^2 d\Omega^2_{(3)} \;, 
\end{equation*}
where 
\begin{equation}
\label{Wrhotau}
W = 1 - \frac{2c}{r^2}  = e^{-4c\tau}\;.
\end{equation}
The effective one-dimensional Lagrangian for spherically symmetric backgrounds is given by
\begin{equation}
	{\cal L}_1 = \tfrac{1}{4}a_{IJ}(\sigma) \left( \dot{\sigma}^I \dot{\sigma}^J - \dot{b}^I \dot{b}^J \right) \;,
\label{1Daction}
\end{equation}
which must be supplemented by the Hamiltonian constraint
\begin{equation}
\tfrac{1}{4} a_{IJ}(\sigma) \left(\dot{\sigma}^I \dot{\sigma}^J - \dot{b}^I \dot{b}^J \right) = c^2 \;. 
\label{on_shell_A2}
\end{equation}

The equations of motion for the one-dimensional Lagrangian (\ref{1Daction}) are
\begin{align*}
\frac{d}{d\tau} \left( a_{IJ}(\sigma) \dot{\sigma}^J \right) - \tfrac{1}{2} \partial_I a_{JK}(\sigma) \left( \dot{\sigma}^J \dot{\sigma}^K - \dot{b}^J \dot{b}^K \right) &= 0 \;, \\
\frac{d}{d\tau} \left( a_{IJ}(\sigma) \dot{b}^J \right) &= 0 \;.
\end{align*}
The equations of motion for $b^I$ can be solved immediately
\[
	a_{IJ}(\sigma) \dot{b}^J = Q_I \;,
\]
where the $Q_I$ are constant electric charges that correspond to the isometry $b^I \to b^I + C^I$. 

The remaining second order equation of motion for $\sigma^I$ becomes much simpler if one introduces a natural set of dual coordinates $\sigma_I$, defined by
\[
	\sigma_I := \partial_I \tilde{H} = -a_{IJ}(\sigma) \sigma^J \;.
\]
It is clear that both coordinates $\sigma^I$ and dual coordinates $\sigma_I$ are related algebraically.
The derivative of $\sigma_I$ can by written using the chain rule as
\[
	\dot{\sigma}_I = \frac{d}{d\tau} \sigma_I = a_{IJ}(\sigma) \dot{\sigma}^J \;.
\]
Plugging the dual coordinates into the second order equations of motion and Hamiltonian constraint we find
\begin{align}
\ddot{\sigma}_I + \tfrac{1}{2} \partial_I a^{JK}(\sigma) \left( \dot{\sigma}_J \dot{\sigma}_K - Q_J Q_K \right) &= 0  \;, \label{eom1} \\
\tfrac{1}{4} a^{IJ}(\sigma) \left(\dot{\sigma}_I \dot{\sigma}_J - Q_I Q_J \right) &= c^2 \;. \label{eom2}
\end{align}
We are left to solve these equations of motion.

Extremal instanton solutions correspond to the choice $c = 0$. In this case the equations of motion can be solved for arbitrary models by\footnote{These solutions necessarily lift to BPS black holes. If the metric of the target manifolds allows for a field rotation matrix $R^I_{\;\;K}$ that satisfies $a_{IJ}R^I_{\;\;K} R^J_{\;\;L} = a_{KL}$ then one can generalise this ansatz to produce solutions which lift to non-BPS black holes \cite{Ceresole:2007wx, Lopes Cardoso:2007ky, Mohaupt:2009iq}.}
\[
	\dot{\sigma}_I = \pm Q_I \;, \qquad \Rightarrow \qquad \sigma_I = A_I \pm Q_I \tau \;. 
\]
Note that the number of possible independent integration constants from $(n+1)$ second order differential equations should be $(2n+2)$, but in the extremal solution above we only have $(n+1)$ integration constants. This is because extremal solutions must satisfy the first order attractor equations, of which much has already been explained in the literature, see for example \cite{Andrianopoli:2007gt, Mohaupt:2009iq}.

We will now investigate non-extremal solutions where $c \neq 0$. This turns out to be considerably more difficult, as the the non-extremality parameter entangles the second order equations of motion in a highly non-trivial manner, and we can only find the most general solution for specific models.


\section{Instanton solutions}

\subsection{General solution of STU-like models}

Let us fix that we have $n$ physical scalar fields $\phi^x$ and a generalised projective special real target manifold. We will consider STU-like models, where the Hesse potential on the corresponding $d$-conical affine special real manifold takes the form
\[
	H(h) = \left( h^0 h^1 \ldots h^n \right)^{\frac{d}{(n+1)}} \;,
\]
or models that can be brought to this form by a linear transformation. We will only consider patches where $h^I$ are pointwise non-zero, and note that by construction the Hesse potential is strictly positive. This class of models actually generalises the class of models for which solutions were found in \cite{Mohaupt:2010fk}, where only the special case $d = (n+1)$ was considered. The supergravity STU model is given by the special case $n = 2$ and $d = 3$. Using the formula (\ref{Phiparam}) the hypersurface $H = 1$ can be parametrised by the $n$ physical scalar fields $\phi^x$ through
\begin{equation}
	\phi^x = \frac{h^x}{h^0} \;, \qquad h^0 = (\phi^1 \ldots \phi^n)^{-\frac{1}{(n+1)}} \;.
	\label{PhysScal}
\end{equation}

We now need to calculate the equations of motion (\ref{eom1}) and (\ref{eom2}) for this class of models. The matrix $a^{IJ}$ and its derivative can be calculated using (\ref{a}), and are given in terms of dual coordinates $\sigma_I$ by
\begin{align*}
	a^{IJ} &= \text{diag}\left( \frac{1}{(n+1)\,\sigma_0^2}, \ldots, \frac{1}{(n+1)\,\sigma_n^2} \right) \;,  \\
	\partial_I a^{JK} &= \text{diag}\left( -\frac{2}{\sigma_0}, \ldots, -\frac{2}{\sigma_n} \right) \;.
\end{align*}
The equations of motion then take the form
\begin{align}
\ddot{\sigma}_I - \frac{\left[(\dot{\sigma}_I)^2 - (Q_I)^2 \right]}{\sigma_I} &= 0  \;,  \label{STUeom1}\\
 \sum_I \frac{\left[(\dot{\sigma}_I)^2 - (Q_I)^2 \right]}{(n+1)\,\sigma_I^2} &= 4c^2 \;. \label{STUeom2}
\end{align}
The second order equations (\ref{STUeom1}) for each coordinate $\sigma_I$ completely decouple from one-another, and can be explicitly integrated to find the general solution
\begin{equation}
	\sigma_I = \pm\frac{Q_I}{B_I} \sinh \left( B_I \tau + B_I \frac{A_I}{Q_I} \right) \;.
	\label{STUgensol}
\end{equation}
The constraint (\ref{STUeom2}) then relates the integration constants with the non-extremality parameter
\begin{equation}
	\frac{1}{(n+1)}(B_0)^2 + \ldots + \frac{1}{(n+1)}(B_n)^2 = 4c^2 \;.
	\label{STUBconst}
\end{equation}
One can either interpret $c$ as a dependent parameter, or see this as a restriction on the integration constants. Either way, after solving all equations of motion we are left with $(2n + 2)$ free parameters. Since the solution is invariant under $B_I \to -B_I$ we can assume without loss of generality that the $B_I$ are non-negative. The Kaluza-Klein scalar can be written in terms of the dual coordinates as
\[
	e^{-\tilde{\sigma}} = (-1)^{(n+1)}(n+1)(\sigma_0 \ldots \sigma_n)^{\frac{1}{(n+1)}} \;.
\]
Note that upon setting $c \to 0$ we immediately have $B_I \to 0$ due to (\ref{STUBconst}). The general solution then reduces to the extremal solution.

\subsection{Block diagonal models}

For models in which the matrix $a_{IJ}$ splits into distinct blocks, or can be made to do so be a linear transformation, we find a restricted class of solutions in which the number of independent scalar fields is the same as the number of blocks. Solutions to each block are given again by the general solution (\ref{STUgensol}). We will demonstrate this with an example that has two blocks.

Consider a model with $n$ physical scalar fields and a generalised projective special real target manifold with a corresponding Hesse potential that is homogeneous of degree $d$. For a general Hesse potential the physical scalar fields can be written using (\ref{Phiparam}) as
\[
	\phi^x = \frac{h^x}{h^0} \;, \qquad h^0 = \hat{H}(\phi^1, \ldots, \phi^n)^{-\frac{1}{d}} := H\left(1, \frac{h^1}{h^0}, \ldots, \frac{h^n}{h^0}\right)^{-\frac{1}{d}} \;.
\]
We will assume that the metric $a_{IJ}$ decomposes into precisely two blocks
\[
	a_{IJ} = \left( \begin{array}{cccccc} 
		* 			& \ldots 	& * 				& 0 & 0 & 0 \\ 
		\vdots 	& \ddots 	& \vdots 		& 0 & 0 & 0 \\ 
		* 			& \ldots 	& * 				& 0 & 0 & 0 \\ 
		0 			& 0 			& 0 				& * 			& \ldots & * \\ 
		0			 	& 0 			& 0				 	& \vdots 	& \ddots & \vdots \\ 
		0 			& 0 			& 0 				& * 			& \ldots & * 
		\end{array}\right) \;.
\]
Let us denote the size of the first block by $k\times k$ and the second block by $l\times l$, so that $k + l = (n+1)$. A Hesse potential that produces such a block diagonal metric is given by
\[
	H(\sigma^0, \ldots, \sigma^n) = H^1(\sigma^0, \ldots, \sigma^{k-1}) H^2(\sigma^k, \ldots, \sigma^n) \;.
\]
We now set all scalar fields within each block to be proportional to one another
\begin{align*}
	\sigma^0 \propto \ldots \propto \sigma^{k-1} \;, \qquad \sigma^k \propto \ldots \propto \sigma^{n} \;,
\end{align*}
which implies that the dual coordinates $\sigma_I$ are proportional to one-another
\begin{align*}
	\sigma_{(0)} := \sigma_0 \propto \ldots \propto \sigma_{k-1} \;, \qquad	\sigma_{(1)} := \sigma_{k} \propto \ldots \propto \sigma_{n} \;.
\end{align*}
The solution is characterised by just two independent scalar fields $\sigma_{(0)}$ and $\sigma_{(1)}$ and two electric charges $Q_{(0)}$ and $Q_{(1)}$, where
\begin{align*}
 Q_{(0)} &:= Q_0 = \frac{\sigma_{1}}{\sigma_0} Q_1 = \ldots = \frac{\sigma_{k-1}}{\sigma_0} Q_{k-1} \;, \\
 Q_{(1)} &:= Q_k = \frac{\sigma_{k+1}}{\sigma_{k}} Q_{k} = \ldots = \frac{\sigma_{n}}{\sigma_{k}} Q_{n} \;.
\end{align*} 
There is only one independent physical scalar field
\begin{align*}
	\phi^{(1)} &:= \phi^k 
												= \frac{\sigma^{k}}{\sigma^{k+1}} \phi^{k+1} = \ldots = \frac{\sigma^{k}}{\sigma^{n}} \phi^n \;,
\end{align*}
and the other physical scalars are constant
\[
	\phi^1 
					= \frac{\sigma^1}{\sigma^2} \phi^2 = \ldots = \frac{\sigma^{1}}{\sigma^{k-1}} \phi^{k -1} = \text{const} \;.
\]
The equations of motion reduce to
\begin{align}
\ddot{\sigma}_{(0)} - \frac{\left[(\dot{\sigma}_{(0)})^2 - (Q_{(0)})^2 \right]}{\sigma_{(0)}} &= 0  \;,  \label{Blockeom1}\\
\ddot{\sigma}_{(1)} - \frac{\left[(\dot{\sigma}_{(1)})^2 - (Q_{(1)})^2 \right]}{\sigma_{(1)}} &= 0  \;,  \label{Blockeom2}\\
 \psi_0 \frac{\left[(\dot{\sigma}_{(0)})^2 - (Q_{(0)})^2 \right]}{\sigma_{(0)}^2} 
 + \psi_1 \frac{\left[(\dot{\sigma}_{(1)})^2 - (Q_{(1)})^2 \right]}{\sigma_{(1)}^2} &= 4c^2 \;, \label{Blockeom3}
\end{align}
where $\psi_0, \psi_1$ are fixed constants that depend on the ratios $\frac{\sigma_x}{\sigma_0}$, and from (\ref{SigId}) they must satisfy the identity
\[
	\psi_0 + \psi_1 = 1 \;.
\] 
Just as for STU-like models, we can find the general solution to the second order equations (\ref{Blockeom1}), (\ref{Blockeom2}), which is given by
\begin{align}
	\sigma_{(0)} &= \pm\frac{Q_{(0)}}{B_{(0)}} \sinh \left(B_{(0)} \tau + B_{(0)} \frac{A_{(0)}}{Q_{(0)}}\right) \;, \label{Blockgensol1} \\
	\sigma_{(1)} &= \pm\frac{Q_{(1)}}{B_{(1)}} \sinh \left(B_{(1)} \tau + B_{(1)} \frac{A_{(1)}}{Q_{(1)}}\right) \;, \label{Blockgensol2}
\end{align}
and the constraint (\ref{Blockeom3}) places one restriction on the integration constants
\begin{equation}
	\psi_0 \left(B_{(0)}\right)^2 + \psi_1 \left(B_{(1)}\right)^2 = 4c^2 \;.
	\label{Blockconst}
\end{equation}
The solution naturally generalises to models with more than two blocks. For a metric with two blocks we obtained solutions characterised by one non-constant scalar field. With three blocks solutions will be characterised by two independent non-constant scalar fields, etc. We can write the Kaluza-Klein scalar as
\[
	e^{-\tilde{\sigma}} = \mu \left( \sigma_{(0)} \right)^{\frac{k}{(n+1)}} \left( \sigma_{(1)} \right)^{\frac{l}{(n+1)}} \;,
\]
where $\mu$ is a fixed constant that depends on the ratios $\frac{\sigma_x}{\sigma_0}$.

Since every matrix can be thought of as having one block (the whole matrix), this method provides at least one universal instanton solution for any model. In this case all the physical scalar fields are constant. We will see in the next section that when we lift the universal solution to five dimensions we obtain the Reissner-Nordstr\"om black hole.


\section{Non-extremal black hole solutions}

The four-dimensional instanton solutions in the previous section can be lifted to static solutions of the five dimensional theory by retracing the steps of dimensional reduction. However, for these solutions to correspond to black holes they need to satisfy certain criteria:
\begin{enumerate}
	\item An event horizon with finite area must exist.
	\item The physical scalar fields $\phi^x$ must take finite values on the horizon.
\end{enumerate}
We will show that these two requirement force us to make restrictions on the integration constants that reduce the number by exactly half - just like the extremal case - which suggests a first order rewriting. The fact that certain non-example black holes are governed by first order equations has been known in the literature for some time \cite{Lu:2003iv, Miller:2006ay, Garousi:2007zb, Andrianopoli:2007gt, Janssen:2007rc, Cardoso:2008gm, Perz:2008kh, Chemissany:2009hq}. But here we present the argument differently. For STU-like models we start with the most general solution to the equations of motion truncated to static, spherically symmetric and purely electric backgrounds. We then impose the above criteria on the general solution, and by doing so find the most general type of non-extremal black hole solution using the parametrisation of the physical scalar fields given by (\ref{Phiparam}). The fact that the number of integration constants reduces by half is interesting because there is no reason a priori that non-extremal solutions should be governed by first order equations. Since we see no reason why the STU-like models should be a privileged with respect to the number of integration constants, it is reasonable to suspect that this is a feature of non-extremal black hole solutions to all models.

\subsection{STU-like models}

We can lift the instanton solutions found in the previous section to a static solution to the five-dimensional theory
\begin{align*}
	ds_5^2 &= -\frac{1}{ (n+1)^2\left( \sigma_0 \ldots \sigma_n \right)^{\frac{2}{(n+1)}} } dt^2 \\
	&\hspace{1em} + (-1)^{(n+1)} (n+1) \left( \sigma_0 \ldots \sigma_n \right)^{\frac{1}{(n+1)}} \left( \frac{c^3}{\sinh^3 2c\tau} d\tau^2 + \frac{c}{\sinh 2c\tau} d\Omega^2_{(3)} \right) \;,
\end{align*}
where one should note that $(-1)^{(n+1)}(n+1)(\sigma_0\ldots \sigma_n)^{\frac{1}{(n+1)}}$ is positive between radial infinity and the outer horizon $\tau \in (0, +\infty)$.
The area $A$ of the outer event horizon is given by
\[
	A = \lim_{\tau \to +\infty} (-1)^{(n+1)} (n+1)\left( \sigma_0 \ldots \sigma_n \right)^{\frac{1}{(n+1)}} \frac{c}{\sinh 2c\tau} \;.
\]
The highest order term in the numerator is proportional to $e^{\frac{1}{(n+1)}(B_0 + \ldots + B_n)\tau}$ (recall that the $B_I$ are non-negative), which must exactly cancel with the highest order term in the denominator $e^{2c\tau}$. We can conclude that in order to obtain a finite area we must have
\begin{equation}
	\frac{1}{(n+1)} \left(B_0 + \ldots + B_n \right) = 2c \;.
	\label{STUint1}
\end{equation}
Next, we turn our attention to the physical scalar fields $\phi^x$. These can be written in terms of the dual scalars $\sigma_I$ simple by
\[
	\phi^x = \frac{\sigma_0}{\sigma_x} \;. 
\]
In the limit $\tau \to +\infty$ the physical scalars $\phi^x$ will not take finite values\footnote{By finite values we mean $\phi^x \longrightarrow \hspace{-1.4em} / \hspace{1.2em}  0, \pm \infty$.} for generic choices of $B_I$. The only way to ensure that they take finite values is to impose
	\[
		B_0 = B_1 = \ldots = B_n \;.
	\]
Combining this with (\ref{STUint1}) we conclude that in order to have a finite horizon and finite scalar fields the integration constants must satisfy
	\begin{equation}
		B_0 = \ldots = B_n = 2c \;.
		\label{STUBrestriction}
	\end{equation}
The solution (\ref{STUgensol}) therefore reduces to 
	\begin{equation}
		\sigma_I = \pm \frac{Q_I}{2c} \sinh\left(2c \tau + 2c\frac{A_I}{Q_I} \right) \;.	
		\label{STUsol}
	\end{equation}
Lastly, in order for the solution to be Minkowski space at radial infinity it must satisfy $e^{\tilde{\sigma}} \to 1$, which places one further constraint on the integration constants
	\begin{equation}
		(-1)^{(n+1)}(n+1) \left[\pm\frac{Q_0}{2c} \sinh\left(2c\frac{A_0}{Q_0} \right) \ldots \pm\frac{Q_n}{2c} \sinh\left(2c\frac{A_n}{Q_n} \right)\right]^{\frac{1}{(n+1)}} = 1 \;.
	\label{STUArestriction}
	\end{equation}
Due to the constraints (\ref{STUBrestriction}) and (\ref{STUArestriction}) the number of integration constants reduces by precisely one half, from $(2n + 2)$ to $(n + 1)$. This is suggestive of a first order rewriting, and indeed this can be achieved by first defining the generating function ${\cal W} = {\cal W}(\sigma^I, Q_I, c)$ by
\[
	{\cal W} := \pm \frac{1}{(n+1)} \sum_I \left[ \sqrt{4c^2 + (n+1)^2 Q_I^2{\sigma^I}^2} + c\log \left(\frac{\sqrt{4c^2 + (n+1)^2 Q_I^2{\sigma^I}^2} - 2c}{\sqrt{4c^2 + (n+1)^2 Q_I^2{\sigma^I}^2} + 2c} \right) \right] \;.
\]
This is of a similar form to the generating function for the four-dimensional STU model \cite{Galli:2011fq}.
We can therefore write the solution as first order flow equations 
\begin{align*}
	\dot{\sigma}_I &= \frac{\partial}{\partial \sigma^I} {\cal W} \;, \\
		&= \pm \sqrt{Q_I^2 + 4c^2 \sigma_I^2} \;,
\end{align*}
which is clearly solved by (\ref{STUsol}). These are first order differential equations (in $\tau$), which relate $\dot{\sigma}_I$ to the gradient of a function. They can alternatively be written as $\dot{\sigma}^I = a^{IJ} \partial_J {\cal W}$.

Collecting everything together, we find that the most general static black hole solution for the STU model is given by
\[
	ds_5^2 = -\frac{W}{ \left( {\cal H}_0 \ldots {\cal H}_n \right)^{\frac{2}{(n+1)}} } dt^2 + \left( {\cal H}_0 \ldots {\cal H}_n \right)^{\frac{1}{(n+1)}} \left( \frac{dr^2}{W} + r^2 d\Omega^2_{(3)} \right) \;,
\]
where 
\begin{align*}
	W &= 1 - \frac{2c}{r^2} \;, & {\cal H}_I &=  \mp (n+1) \left[ \frac{Q_I}{2c} \sinh \left(2c\frac{A_I}{Q_I}\right) + \frac{Q_I e^{-2c\frac{A_I}{Q_I}}}{2} \frac{1}{r^2} \right] \;, \\
	&= e^{-4c\tau} \;, & &= \mp (n+1) \left[\frac{1}{4c}Q_I e^{2c\frac{A_I}{Q_I}} -   \frac{1}{4c}Q_I e^{-2c\frac{A_I}{Q_I}} e^{-4c\tau}\right] \;,
\end{align*}
and the scalar fields are given by
\begin{align*}
	\phi^x =	\frac{\sigma_0}{\sigma_x} \;, \qquad \sigma_I &= \frac{-1}{(n+1)}\frac{{\cal H}_I}{\sqrt{W}} = \pm\frac{Q_I}{2c} \sinh\left(2c\tau + 2c\frac{A_I}{Q_I}\right) \;. 
\end{align*}
For the case where $n = 2$ and $d = 3$ this reproduces the non-extremal black hole solutions of $5D, \, {\cal N} = 2$ supergravity originally found in \cite{Maldacena:1996ix, Maldacena:1996ky}.

\subsection{Block diagonal models}

Let us now lift the instanton solutions to models with block diagonal matrix $a_{IJ}$, described in the previous section, to static solutions in five dimensions. Again we will focus on an example with two blocks of size $k\times k$ and $l\times l$. The line element is given by
\begin{align*}
	ds_5^2 &= -\frac{1}{ \mu \left( \sigma_{(0)} \right)^{\frac{2k}{(n+1)}} \left( \sigma_{(1)} \right)^{\frac{2l}{(n+1)}} } dt^2 \\
	&\hspace{4em} +  \mu \left( \sigma_{(0)} \right)^{\frac{k}{(n+1)}} \left( \sigma_{(1)} \right)^{\frac{l}{(n+1)}} \left( \frac{c^3}{\sinh^3 2c\tau} d\tau^2 + \frac{c}{\sinh 2c\tau} d\Omega^2_{(3)} \right) \;.
\end{align*}
The area $A$ of the outer event horizon is given by
\[
	A = \lim_{\tau \to +\infty} \mu \left( \sigma_{(0)} \right)^{\frac{k}{(n+1)}} \left( \sigma_{(1)} \right)^{\frac{l}{(n+1)}} \frac{c}{\sinh 2c\tau} \;.
\]
The highest order term in the numerator is proportional to $e^{\left(\frac{k}{(n+1)}B_{(0)} + \frac{l}{(n+1)}B_{(1)}\right) \tau}$, which must exactly cancel with the highest order term in the denominator $e^{2c\tau}$. We can conclude that in order to obtain a finite area we must have
\begin{equation*}
	\frac{k}{(n+1)} B_{(0)} + \frac{l}{(n+1)} B_{(1)} = 2c \;.
	\label{Blockint1}
\end{equation*}
The physical scalar field $\phi^{(1)}$ can be written in terms of the dual scalars $\sigma_{(0,1)}$ as
\[
	\phi^{(1)} \sim \frac{\sigma_{(0)}}{\sigma_{(1)}} \;.
\]
In the limit $\tau \to +\infty$ the physical scalar $\phi^{(1)}$ will not take finite values for generic choices of $B_{(0,1)}$. The only way to ensure that they take finite values is to impose
\[
	B_{(0)} = B_{(1)} \;.
\]
Combining this with (\ref{Blockint1}) we conclude that in order to have a finite horizon and finite scalar fields the integration constants must satisfy
\[
	B_{(0)} = B_{(1)} = 2c \;.
\]
Ensuring that the solution is Minkowski space at radial infinity $e^{\tilde{\sigma}} \to 1$ places one further constraint on the integration constants
	\begin{equation*}
		\mu \left(\pm \frac{Q_{(0)}}{2c} \sinh\left(2c \frac{A_{(0)}}{Q_{(0)}}\right) \right)^{\frac{k}{(n+1)}} \left(\pm \frac{Q_{(1)}}{2c} \sinh\left(2c \frac{A_{(1)}}{Q_{(1)}}\right) \right)^{\frac{l}{(n+1)}} = 1 \;.
	\end{equation*}

Collecting everything together, we find that our solution for the block diagonal models can be written as
\[
	ds_5^2 = -\frac{W}{ \left( {\cal H}_{(0)} \right)^{\frac{2k}{(n+1)}} \left( {\cal H}_{(1)} \right)^{\frac{2l}{(n+1)}} } dt^2 + \left( {\cal H}_{(0)} \right)^{\frac{k}{(n+1)}} \left( {\cal H}_{(1)} \right)^{\frac{l}{(n+1)}} \left( \frac{dr^2}{W} + r^2 d\Omega^2_{(3)} \right) \;,
\]
where 
\begin{align*}
	W &= 1 - \frac{2c}{r^2}	= e^{-4c\tau} \;, \\
	{\cal H}_{(0,1)} &= \pm \mu \left[  \frac{Q_{(0,1)}}{2c} \sinh \left(2c\frac{A_{(0,1)}}{Q_{(0,1)}}\right) +  \frac{Q_{(0,1)} e^{-2c\frac{A_{(0,1)}}{Q_{(0,1)}}}}{2} \frac{1}{r^2} \right] \;, \\
 &= \pm \mu \left[ \frac{1}{4c}Q_{(1,2)} e^{2c\frac{A_{(0,1)}}{Q_{(0,1)}}} -   \frac{1}{4c}Q_{(0,1)} e^{-2c\frac{A_{(0,1)}}{Q_{(0,1)}}} e^{-4c\tau} \right] \;.
\end{align*}
and the scalar fields are given by
\begin{align*}
	\phi^{{(1)}} &\sim	\frac{\sigma_{(0)}}{\sigma_{{(1)}}} \;, \notag \\
	\sigma_{{(0,1)}} &= \frac{1}{\mu} \frac{{\cal H}_{{(0,1)}}}{\sqrt{W}} = \pm\frac{Q_{{(0,1)}}}{2c} \sinh \left(2c\tau + 2c\frac{A_{(0,1)}}{Q_{(0,1)}}\right) \;.	\end{align*}

\section{Conclusion and Outlook}

We have discussed the notion of a $d$-conical affine special real manifolds and correspondingly generalised projective special real manifolds. The latter generalises the geometry of projective special real manifolds, which appear as the target manifolds of $5D, \, {\cal N} = 2$ supergravity coupled to vector multiplets. We used this to construct a class of five-dimensional gravity-scalar-vector theories that generalises ${\cal N} = 2$ supergravity coupled to vector multiplets. 

Through dimensional reduction and the specific field redefinition (\ref{FieldRedef}) one can obtain a particularly simple effective Lagrangian for static, spherically symmetric and purely electric solutions (\ref{4dLag}). One key feature was that we worked always at the level of the `larger' moduli space: the $d$-conical affine special real manifold. We then focused on STU-like models, where we found the general solution to the equations of motion, and models that are block diagonal, where we found solution with as many independent scalar fields as there are blocks. Since the metrics of all models contain at least one block, this also provides a universal solution to all models. 
We then investigated which solutions correspond to non-extremal black holes solutions of the five-dimensional theory. In order to obtain a finite horizon area and finite scalar fields the number of integration constants must halve, suggesting a first order rewriting of the equations of motion. For STU-like models all calculations were performed explicitly, and at every stage we can set $c \to 0$ to obtain the extremal solution. Since we see no reason STU-like models should be privileged in their number of integration constants, we conjecture that all non-extremal black hole solutions should have half the number of integration constants one would expect from the second order equations of motion.

One obvious extension to this work is to investigate solutions of more complicated models. However, it was shown in \cite{Mohaupt:2010fk} that the hyperbolic-sine form of the solution to STU-like models (\ref{STUsol}) does not give the most general solution for generic models. 
One must therefore replace the hyperbolic-sine function with something more complicated. It is an open question whether one can find a general formula for such a function, e.g. \cite{Meessen:2011bd}, or whether one can only find explicit formulas for specific models. At this point it is still not even clear in the literature that every extremal black hole solution admits a non-extremal generalisation \cite{Bossard:2009at, Chemissany:2009hq}. 

One may also wonder whether this analysis can be repeated for four-dimensional theories. In \cite{Mohaupt:2011aa} it was shown that the effective action for static solutions to $4D, \, {\cal N} = 2$ supergravity coupled to vector multiplets can be brought to the same simple form as (\ref{4dLag}) for general static spacetime metrics (see p51 of \cite{Mohaupt:2011aa}). One can then follow exactly the same logic for axion-free solutions to the four-dimensional STU model as we have present here for the five-dimensional STU model: one can find the general solution to the equations of motion, and show that these correspond to black hole solutions with finite scalar fields only when the number of integration constants reduces by half. This will be presented in future work \cite{Future}.

Another natural extension is to consider various other types of solutions. These include solutions with a cosmological constant or Taub-NUT charge, rotating solutions, domain walls, black strings and cosmological solutions. Reduction over time has previously been used to construct black ring solutions \cite{Yazadjiev:2005pf, Yazadjiev:2006hw, Yazadjiev:2006ew}, and in \cite{Meessen:2012su} black ring solutions were constructed based on \cite{Meessen:2011aa}. Cosmological solutions may also be particularly interesting as the non-extremal black hole solutions we have discussed can be continued beyond the horizon where the Killing vector is spacelike. This provides a natural starting point for the construction of S-brane cosmological solutions \cite{Behrndt:1994ev, Gutperle:2002ai}.

Theories of gauged supergravity are also applicable to the analysis presented in this paper. In \cite{Klemm:2012yg} it was shown that the same procedure can be used to find new non-extremal solutions to four-dimensional Fayet-Iliopoulos gauged supergravity. It would also be interesting to investigate non-extremal solutions of five-dimensional gauged supergravity, though solutions to the STU model have previously been found by other methods \cite{Behrndt:1998jd}.

Lastly, one may wonder whether special K\"ahler geometry, which corresponds to $4D, \, {\cal N} = 2$ supergravity coupled to vector multiplets, can be generalised in a way similar to the generalisation of special real geometry considered in this paper. At present there does not exist a well defined generalisation of special K\"ahler geometry. However, the dimensional reduction of $d$-conical affine special real geometry suggests that generalising the degree of homogeneity of the holomorphic prepotential may provide one consistent generalisation of conic affine special K\"ahler geometry. This would be interesting to investigate in the future.

\subsubsection*{Acknowledgments}

The work of T.M. is supported in part by STFC  
grant ST/G00062X/1. The work of O.V. is supported by an STFC studentship and DAAD.




\begin{thebibliography}{99}

\bibitem{Ferrara:1995ih}
	S. Ferrara, R. Kallosh and A. Strominger,
	{\em N=2 extremal black holes},
	Phys. Rev. D 52 (1995) 5412,
	hep-th/9508072.

\bibitem{Strominger:1996sh}
	A. Strominger and C. Vafa,
	{\em Microscopic Origin of the Bekenstein-Hawking Entropy},
	Phys. Lett. B 379 (1996) 99,
	hep-th/9601029.
	
\bibitem{Maldacena:1997de}
  J.~M.~Maldacena, A.~Strominger and E.~Witten,
  JHEP {\bf 9712} (1997) 002
  [hep-th/9711053].
  
\bibitem{Vafa:1997gr}
  C.~Vafa,
  ``Black holes and Calabi-Yau threefolds,''
  Adv.\ Theor.\ Math.\ Phys.\  {\bf 2} (1998) 207
  [hep-th/9711067].
  
\bibitem{Lopes Cardoso:1998wt}
  G.~Lopes Cardoso, B.~de Wit and T.~Mohaupt,
  ``Corrections to macroscopic supersymmetric black hole entropy,''
  Phys.\ Lett.\ B {\bf 451} (1999) 309
  [hep-th/9812082].

\bibitem{Callan:1996dv}
C.G. Callan and J.M. Maldacena,
{\em D-brane Approach to Black Hole Quantum Mechanics},
Nucl. Phys. B 472 (1996) 591,
hep-th/9602043.

\bibitem{Horowitz:1996fn}
G.T. Horowitz and A. Strominger,
{\em Counting States of Near-Extremal Black Holes},
Phys. Rev. Lett. 77 (1996) 2368,
hep-th/9602051.


\bibitem{Cvetic:1996kv}
M. Cvetic and D. Youm,
{\em Entropy of non-extreme charged rotating black holes in string theory},
Phys. Rev. D 54 (1996) 2612,
hep-th/9603147.

\bibitem{Maldacena:1996ix}
J.M. Maldacena and A. Strominger,
{\em Black hole greybody factors and D-brane spectroscopy},
Phys. Rev. D 55 (1997) 861,
hep-th/9609026. 


\bibitem{Behrndt:1997as}
K. Behrndt, M. Cvetic and W.A. Sabra,
{\em The entropy of near-extreme $N=2$ black holes},
Phys. Rev. D 58 (1998) 084018,
hep-th/9712221.

\bibitem{Cecotti:1988qn}
  S.~Cecotti, S.~Ferrara and L.~Girardello,
  ``Geometry of Type II Superstrings and the Moduli of Superconformal Field Theories,''
  Int.\ J.\ Mod.\ Phys.\ A {\bf 4} (1989) 2475.

\bibitem{Ferrara:1989ik}
  S.~Ferrara and S.~Sabharwal,
  ``Quaternionic Manifolds for Type II Superstring Vacua of Calabi-Yau Spaces,''
  Nucl.\ Phys.\ B {\bf 332} (1990) 317.
  
\bibitem{deWit:1991nm}
  B.~de Wit and A.~Van Proeyen,
  ``Special geometry, cubic polynomials and homogeneous quaternionic spaces,''
  Commun.\ Math.\ Phys.\  {\bf 149} (1992) 307
  [hep-th/9112027].  
  
\bibitem{deWit:1995tf}
  B.~de Wit and A.~Van Proeyen,
  ``Isometries of special manifolds,''
  hep-th/9505097.

\bibitem{Gunaydin:2005mx}
  M.~Gunaydin, A.~Neitzke, B.~Pioline and A.~Waldron,
  ``BPS black holes, quantum attractor flows and automorphic forms,''
  Phys.\ Rev.\ D {\bf 73} (2006) 084019
  [hep-th/0512296].

\bibitem{Bossard:2009at}
  G.~Bossard, H.~Nicolai and K.~S.~Stelle,
  ``Universal BPS structure of stationary supergravity solutions,''
  JHEP {\bf 0907} (2009) 003
  [arXiv:0902.4438 [hep-th]].

\bibitem{Chemissany:2009hq}
  W.~Chemissany, J.~Rosseel, M.~Trigiante and T.~Van Riet,
  ``The Full integration of black hole solutions to symmetric supergravity theories,''
  Nucl.\ Phys.\ B {\bf 830} (2010) 391
  [arXiv:0903.2777 [hep-th]].

\bibitem{Bossard:2009we}
  G.~Bossard, Y.~Michel and B.~Pioline,
  ``Extremal black holes, nilpotent orbits and the true fake superpotential,''
  JHEP {\bf 1001} (2010) 038
  [arXiv:0908.1742 [hep-th]].  
  
\bibitem{AzregAinou:2011gt}
  M.~Azreg-Ainou, G.~Clement and D.~V.~Gal'tsov,
  ``All extremal instantons in Einstein-Maxwell-dilaton-axion theory,''
  Phys.\ Rev.\ D {\bf 84} (2011) 104042
  [arXiv:1107.5746 [hep-th]].

\bibitem{Cortes:2011}
	V.~Cort\'es, J.~Louis, P.~Smyth, H.~Triendl,
	``On certain K\"ahler quotients of quaternionic K\"ahler manifolds,''
	arXiv:1111.0679 [math.DG].
  
\bibitem{Mohaupt:2011aa}
  T.~Mohaupt and O.~Vaughan,
  ``The Hesse potential, the c-map and black hole solutions,''
  JHEP {\bf 1207} (2012) 163
  [arXiv:1112.2876 [hep-th]].
  
\bibitem{Alexandrov:2012bn}
  S.~Alexandrov,
  ``c-map as c=1 string,''
  Nucl.\ Phys.\ B {\bf 863} (2012) 329
  [arXiv:1201.4392 [hep-th]].

\bibitem{Alekseevsky:2012}
	D.~V.~Alekseevsky, V.~Cort{\'e}s and T.~Mohaupt,
	``Conification of K\"ahler and hyper-K\"ahler manifolds,''
	arXiv:1205.2964 [math.DG]. 

\bibitem{Yazadjiev:2005pf}
  S.~S.~Yazadjiev,
  ``Generating dyonic solutions in 5D Einstein-dilaton gravity with antisymmetric forms and dyonic black rings,''
  Phys.\ Rev.\ D {\bf 73} (2006) 124032
  [hep-th/0512229].
  
\bibitem{Gaiotto:2007ag}
  D.~Gaiotto, W.~Li and M.~Padi,
  ``Non-Supersymmetric Attractor Flow in Symmetric Spaces,''
  JHEP {\bf 0712} (2007) 093
  [arXiv:0710.1638 [hep-th]].

\bibitem{Janssen:2007rc}
	B. Janssen, P. Smyth, T. Van Riet and B. Vercnocke,
	{\em A first-order formalism for timelike and spacelike brane
	solutions},
	JHEP 04 (2008) 007,
	arXiv:0712.2808.

\bibitem{Perz:2008kh}
	J.~Perz, P.~Smyth, T.~Van Riet and B.~Vercnocke,
	{\em First-order flow equations for extremal and non-extremal black holes},
	JHEP {\bf 0903} (2009) 150
	arXiv:0810.1528.

\bibitem{Breitenlohner:1987dg}
	P. Breitenlohner, D. Maison and G.W. Gibbons,
	{\em Four-Dimensional Black Holes from Kaluza-Klein Theories},
	Commun. Math. Phys. 120 (1988) 295.
	
\bibitem{Mohaupt:2010fk}
  T.~Mohaupt and O.~Vaughan,
  ``Non-extremal Black Holes, Harmonic Functions, and Attractor Equations,''
  Class.\ Quant.\ Grav.\  {\bf 27} (2010) 235008
  [arXiv:1006.3439 [hep-th]].

\bibitem{Mohaupt:2009iq}
  T.~Mohaupt and K.~Waite,
  ``Instantons, black holes and harmonic functions,''
  JHEP {\bf 0910} (2009) 058
  [arXiv:0906.3451 [hep-th]].

\bibitem{Lu:2003iv}
H. Lu, C.N. Pope and J.F. Vazquez-Poritz,
{\em From AdS black holes to supersymmetric flux branes},
Nucl. Phys. B 709 (2005) 47,
hep-th/0307001.

\bibitem{Miller:2006ay}
C.M. Miller, K. Schalm and E.J. Weinberg,
{\em Nonextremal black holes are BPS},
Phys. Rev. D 76 (2007) 044001,
hep-th/0612308.


\bibitem{Garousi:2007zb}
M.R. Garousi and A. Ghodsi,
{\em On attractor mechanism and entropy function for
non-extremal black holes},
JHEP 05 (2007) 043,
hep-th/0703260.

\bibitem{Andrianopoli:2007gt}
L. Andrianopoli, R. D'Auria and E. Orazi,
{\em First order description of black holes in moduli space},
JHEP 11 (2007) 032,
arXiv:0706.0712.

\bibitem{Cardoso:2008gm}
G.L. Cardoso and V. Grass,
{\em On five-dimensional non-extremal charged black holes and 
FRW cosmology},
Nucl. Phys. B 803 (2008) 209,
arXiv:0803.2819.

\bibitem{Gunaydin:1983bi}
  M.~Gunaydin, G.~Sierra and P.~K.~Townsend,
  ``The Geometry of N=2 Maxwell-Einstein Supergravity and Jordan Algebras,''
  Nucl.\ Phys.\ B {\bf 242} (1984) 244.

\bibitem{Cortes:2011aj}
  V.~Cortes, T.~Mohaupt and H.~Xu,
  ``Completeness in supergravity constructions,''
  arXiv:1101.5103 [hep-th].

\bibitem{AlekseevskyRigid}
  D.~V.~Alekseevsky, V.~Cortes,
  ``Geometric construction of the r-map: from affine special real to special K\"ahler manifolds,''
  Comm.\ Math.\ Phys.\ 291 (2009), 579-590 
  [arXiv:0811.1658 [math.DG]].
  
\bibitem{Mohaupt:2011ab}
  T.~Mohaupt and O.~Vaughan,
  ``Developments in special geometry,''
  J.\ Phys.\ Conf.\ Ser.\  {\bf 343} (2012) 012078
  [arXiv:1112.2873 [hep-th]].

\bibitem{Gibbons:1998xa}
  G.~W.~Gibbons and P.~Rychenkova,
  ``Cones, triSasakian structures and superconformal invariance,''
  Phys.\ Lett.\ B {\bf 443} (1998) 138
  [hep-th/9809158].

\bibitem{Meessen:2011bd}
  P.~Meessen and T.~Ortin,
  ``Non-Extremal Black Holes of N=2,d=5 Supergravity,''
  Phys.\ Lett.\ B {\bf 707} (2012) 178
  [arXiv:1107.5454 [hep-th]].
  
    
\bibitem{Ceresole:2007wx}
  A.~Ceresole and G.~Dall'Agata,
  ``Flow Equations for Non-BPS Extremal Black Holes,''
  JHEP {\bf 0703} (2007) 110
  [hep-th/0702088].

\bibitem{Lopes Cardoso:2007ky}
  G.~Lopes Cardoso, A.~Ceresole, G.~Dall'Agata, J.~M.~Oberreuter and J.~Perz,
  ``First-order flow equations for extremal black holes in very special geometry,''
  JHEP {\bf 0710} (2007) 063
  [arXiv:0706.3373 [hep-th]].

\bibitem{Galli:2011fq}
  P.~Galli, T.~Ortin, J.~Perz and C.~S.~Shahbazi,
  ``Non-extremal black holes of N=2, d=4 supergravity,''
  JHEP {\bf 1107} (2011) 041
  [arXiv:1105.3311 [hep-th]].

\bibitem{Maldacena:1996ky}
  J.~M.~Maldacena,
  ``Black holes in string theory,''
  hep-th/9607235.
  
\bibitem{Future}
  T.~Mohaupt and O.~Vaughan,
  to appear.  

\bibitem{Yazadjiev:2006hw}
  S.~S.~Yazadjiev,
  ``Completely integrable sector in 5-D Einstein-Maxwell gravity and derivation of the dipole black ring solutions,''
  Phys.\ Rev.\ D {\bf 73} (2006) 104007
  [hep-th/0602116].
  
\bibitem{Yazadjiev:2006ew}
  S.~S.~Yazadjiev,
  ``Solution generating in 5D Einstein-Maxwell-dilaton gravity and derivation of dipole black ring solutions,''
  JHEP {\bf 0607} (2006) 036
  [hep-th/0604140].

\bibitem{Meessen:2012su}
  P.~Meessen, T.~Ortin, J.~Perz and C.~S.~Shahbazi,
  ``Black holes and black strings of N=2, d=5 supergravity in the H-FGK formalism,''
  arXiv:1204.0507 [hep-th].
  
\bibitem{Meessen:2011aa}
  P.~Meessen, T.~Ortin, J.~Perz and C.~S.~Shahbazi,
  ``H-FGK formalism for black-hole solutions of N=2, d=4 and d=5 supergravity,''
  Phys.\ Lett.\ B {\bf 709} (2012) 260
  [arXiv:1112.3332 [hep-th]].

\bibitem{Behrndt:1994ev}
  K.~Behrndt and S.~Forste,
 	``String Kaluza-Klein cosmology,''
  Nucl.\ Phys.\ B {\bf 430} (1994) 441
  [hep-th/9403179].
    
\bibitem{Gutperle:2002ai}
  M.~Gutperle and A.~Strominger,
  ``Space - like branes,''
  JHEP {\bf 0204} (2002) 018
  [hep-th/0202210].

\bibitem{Klemm:2012yg}
  D.~Klemm and O.~Vaughan,
  ``Nonextremal black holes in gauged supergravity and the real formulation of special geometry,''
  arXiv:1207.2679 [hep-th].
  
\bibitem{Behrndt:1998jd}
  K.~Behrndt, M.~Cvetic and W.~A.~Sabra,
  ``Nonextreme black holes of five-dimensional N=2 AdS supergravity,''
  Nucl.\ Phys.\ B {\bf 553} (1999) 317
  [hep-th/9810227].
          
\end{thebibliography}
\end{document}